\documentstyle[12pt]{article}
\newcommand{\be}{\begin{equation}}
\newcommand{\ee}{\end{equation}}
\newcommand{\al}{\mbox{$\alpha$}}

\newcommand{\bi}[1]{\bibitem{#1}}
\newcommand{\fr}[2]{\frac{#1}{#2}}

\newcommand{\gm}{\mbox{$\gamma_{\mu}$}}

\newcommand{\ph}{\mbox{$\hat{p}$}}

\newcommand{\qh}{\mbox{$\hat{q}$}}

\newcommand{\FD}{\mbox{$\tilde{F}$}}
\newcommand{\gf}{\mbox{$\gamma_{5}$}}

\newcommand{\Ima}{\mbox{Im}}

\begin{document}
\pagestyle{empty}
\normalsize
\begin{flushright}{BINP 95-87\\November 1995}
\end{flushright}
\vspace{0.5cm}
\begin{center}{\Large \bf CP-odd interaction of quarks and SU(2)
gauge bosons with scalar field background in
Kobayashi-Maskawa model.}\\

\vspace{1.5cm}

{\bf M.E.Pospelov }\footnote{E-mail:pospelov@inp.nsk.su} \\

\vspace{0.5cm}

{\em Budker Institute of Nuclear Physics, 630090 Novosibirsk,
Russia}

\end{center}
\vspace{1.5cm}

\begin{abstract} CP-violating interaction of quarks and
W-bosons with the scalar field background is studied in the
Kobayashi-Maskawa with with Standard Model content of flavors
and with additional heavy generation of fermions. The corresponding
two-loop induced formfactors are calculated at zero temperature.
The results are generalized at large momentum transfers to
take into account CP-violating effects in the Higgs boson decay.
The inclusion of this interaction into the scheme of adiabatic
baryogenesis at the temperature of electroweak phase transition
suffers from the uncertainties come from the poor knowledge of the vacuum
condensate value triggering baryon number violating
processes. It is shown, however, that even in the most favorable
assumptions the Standard Model with four generations
cannot produce enough C and CP violation for
the explanation of the observable excess of baryons in the
Universe.

\end{abstract}
\newpage
\pagestyle{plain}
\pagenumbering{arabic}
\section{Introduction}
The violation of CP symmetry observed 30 years ago \cite{FK} is
still considered being an intriguing question of modern physics.
The Kobayashi-Maskawa (KM) model is now the minimal
explanation of this phenomenon in the framework of the Standard
Model (SM). In this work we shall investigate in details the
CP-odd interaction of the neutral scalar field with quarks and
charged gauge bosons in KM model with the standard number of
flavors and with additional heavy generation.

The search for CP-violating decays of Higgs boson into the pair
of quarks or gauge bosons which may not be hopeless to see at
future colliders has been investigated (theoretically) in
various extensions of SM \cite{Higgs}. The KM model predictions
for this processes was not considered up to now though. It is
{\em a priori} clear that SM predictions for the amplitudes of
interest are really tiny and therefore are of methodological
interest only. The situation could be different when the theory
contains one additional heavy fermion generation which existence
is still allowed by current LEP data. Due to the absence of
decoupling for heavy fermions in electroweak theory their
influence on the low energy sector is significant. This
work is aimed at the study of CP-violating amplitudes involving
Higgs boson in the simplest extension of SM by new heavy
generation of fermions with their dependence of unknown masses
and mixing angles.

The second problem where the question of such interaction
naturally arises is the baryogenesis at the temperature of the
electroweak phase transition \cite{Rub}. The promising feature
of electroweak baryogenesis is in the possibility to explain the
excess of baryons over antibaryons in the Universe without
appealing to the GUT scale and staying in general on the
background of known interactions. It is commonly understood,
however, that the SM CP violation is capable to generate the
asymmetry many orders of magnitude smaller than its
experimentally observed value and this is the strong reason
to look for a new CP-violating physics beyond SM.

The original analysis of the CP violation required for
baryogenesis in KM model with three and four generations was
done by Shaposhnikov \cite{Shap}. Recently this problem has
attracted serious attention again \cite{ShF,Gav,HS} in connection
with the "chiral transport" scenario \cite{CKN} according to which
the baryoproduciton occurs in the vicinity of the narrow domain wall
separating different phases.  The opposite case of thick slowly
moving wall allows for the "adiabatic" treatment of baryogenesis
along the scenario considered first in
Refs.\cite{TZ,DHSS,McLSTV,CKN1}. In this case the preferential
production of baryons is governed by the effective C- and CP-odd
interaction of quarks and $SU(2)$ gauge fields with the slowly
varying vacuum expectation value (v.e.v.) of the scalar field.
The magnitude of this interaction in KM model, being the matter
of independent interest, may serve also as an additional check of
the conclusions obtained in the general case \cite{Gav,HS}.
Models with one or several additional heavy generations deserve
special analysis in this respect.

The organization of the article is following: Sections 2 and
3 contain the calculation of the CP-odd interaction of particles
with Higgs background at zero temperature. Section 4 generalizes
the results at $T\neq 0$. It comprises the estimate of
the effect for the SM case and the calculation of the
corresponding couplings for its four generation extension.
In Appendix we extend the results obtained in
Sections 3 at the case of large momentum transfers to include
the effects of CP violation in the decay of the real Higgs particle.

\section{Flavor symmetry of amplitudes}

It turns out that the two-loop induced CP-odd interaction
discussed in \cite{Shap} is easily calculable in zero
temperature limit. Below we shall demonstrate that the effect is
finite and satisfies all constraints imposed by the flavor
interchange symmetry and V - A nature of charged currents.

The interaction of particles with the scalar field at small
momentum transfer is commonly treated using the low energy
theorem (See, for ex., the textbook \cite{Ok}). It allows to
obtain the amplitude of interest from the two-point self-energy function
responsible for the propagation in the constant scalar field
background $v+\chi$. Let us take, for example, the decay of the Higgs
boson into two photons induced via one loop with heavy
fermion, $m_f\gg m_{Higgs}$. The amplitude for this process
could be easily reproduced as a first term of expansion in
$\chi/v$ of the self-energy operator:
\be
\fr{\al}{12\pi}\log\left(\fr{\Lambda^2}{(v+\chi)^2}\right)
F_{\mu\nu}F^{\mu\nu}\longrightarrow
-\fr{\al}{6\pi}\fr{\chi}{v} F_{\mu\nu}F^{\mu\nu},
\ee
where $\Lambda$ is the ultraviolet cutoff, $\chi$ is the Higgs
boson field, $v$ is the vacuum expectation value (v.e.v.) and
$F_{\mu\nu}$ is the tensor of electromagnetic field. A similar
attempt to apply this logic to the CP-odd interaction of quarks
or gauge bosons with the scalar field background would stimulate
one to look for the $(v+\chi)^2$-dependent part of corresponding
CP-odd operators like $m_q\bar{q}i\gf q$ and $F^a\FD^a$.
This dependence appears together with ultraviolet cutoff
dependence. In SM it happens first in the fourteenth order of
perturbation theory \cite{EG,KhV}. We shall demonstrate, however,
that already two-loop level is sufficient to induce the
interaction of quarks and $SU(2)$ gauge bosons with the neutral
scalar field background.

The CP violation in the model originates from the complexity of
the KM matrix elements. For flavor-conserving amplitudes it manifests
first in the quartic combination of KM matrices. The
corresponding general diagrams are depicted at Fig. 1a and 1b.
The solid line here denotes quark Green functions, wavy lines
correspond to W-bosons. At the moment we choose the unitary
gauge as containing the minimum of possible diagrams. These
graphs are taken in the slow varying background of scalar
field. The result could be presented in the form of the
Effective Lagrangian as a series of operators with increasing
power of space-time derivatives from this scalar field and we
would keep only first non-vanishing terms. Since the Lorentz
structure of the matrix element for the on-shell fermion and
W-boson scattering off the scalar field is fixed:
\be
M= A(q^2)i\bar{q}(p_1)\gf q(p_2)+ B(q^2)\epsilon_{\alpha\beta\mu\nu}
W^*_\alpha (p_1)p_{1\beta}W_\mu(p_2) p_{2\nu},
\label{eq:start}
\ee
we intend to calculate in fact the values of corresponding
formfactors $A(q^2)$ and $B(q^2)$ at zero momentum transfer,
$q^2=(p_2-p_1)^2=0$.  Both these Lorentz structures vanish when
$q_\mu\longrightarrow 0$ and therefore we cannot put $q=0$
from the very beginning. This ruins the naive approach
based on the calculation of the two-point functions. Performing
the calculation in the momentum representation we expand the
amplitude over the small momentum $q_\mu$ of the external Higgs
boson and keep both zeroth and first order terms of the
expansion. Technically this resembles at some points the
analysis of the induced electric dipole moment (EDM) which is
known to vanish to two loops \cite{Shab}.

First we determine the flavor arrangement along the fermion
line. Let us denote by $f$ the Green function of $f$-flavored
fermion. Then a CP-odd amplitude for the quark scattering could
be written in the following form:
\be
\sum_{j,k,l}i\Ima(V^*_{jf}V_{jk}V^*_{lk}V_{lf})\;fjklf.
\label{eq:fs1}
\ee
For the fermionic loop at Fig. 1b the corresponding structure looks as:
\be
\sum_{f,j,k,l}i\Ima(V^*_{jf}V_{jk}V^*_{lk}V_{lf})\;fjkl,
\label{eq:fs2}
\ee
where the cyclic permutation of the kind $fjkl=lfjk=klfj=jklf$ is
allowed. It is easy to see that
independently on the number of families the expressions
(\ref{eq:fs1}) and (\ref{eq:fs2}) are antisymmetric under the
interchange of flavors $j$ and $l$:
\be
\sum_{j,k,l}i\Ima(V^*_{jf}V_{jk}V^*_{lk}V_{lf})\;fjklf=
\fr{1}{2}\sum_{j,k,l}i\Ima(V^*_{jf}V_{jk}V^*_{lk}V_{lf})\;f(jkl-lkj)f.
\ee

The extraction of the CP-odd part of the amplitude is simple
when we deal with Standard Model with a sole imaginary phase. To
be concrete we take a scattering of $u$ quark in the scalar field
background. Then the arrangement of flavors inside the loops is
determined uniquely:
\be
i\tilde{\delta}u\left(d(c-t)s-s(c-t)d+s(c-t)b-b(c-t)s+b(c-t)d-d(c-t)b
\right)u,
\label{eq:u}
\ee
where $\tilde{\delta}=\delta c_1c_2c_3s_1^2s_2s_3$ is the only
possible CP-odd invariant of 3 by 3 KM matrix in standard
parametrization \cite{Ok}.

The specific antisymmetrization of these amplitude in flavors
causes, according to Shabalin \cite{Shab}, the identical
cancellation of diagrams corresponding to electric dipole moments of
quarks. The same is true for the EDM of W-boson and electron
\cite{KhP}. Therefore, we have to find out first wether
the graphs determining the scattering in the scalar
field background survive under the antisymmetrization in flavor.

The "dangerous" block responsible for the vanishing of EDMs
comprises a mass operator (vertex part) between two fermion
Green functions corresponding to different flavors.
Taking into account all possible ways of external Higgs
attachment, Fig. 2, we write down a general expression for this
block:
\begin{eqnarray}
\fr{\chi(q)}{v}\fr{1-\gf}{2}\left[S_j(p-q/2)m_jS_j(p+q/2)M(p+q/2)
S_k(p+q/2)\right.\nonumber\\ +S_j(p-q/2)\Gamma(p,q)S_k(p+q/2) \nonumber\\
\left.+S_j(p-q/2)M(p-q/2)S_k(p-q/2)m_kS_k(p+q/2)\right]\fr{1+\gf}{2}
\;-\;(j\leftrightarrow k),
\label{eq:struct}
\end{eqnarray}
where $S_j(p)=i(\ph-m_j)^{-1}$ is the Green function of the
$j$-flavored quark and $\ph\equiv\gamma^\mu p_\mu$. $M(p)$ and
$\Gamma(p,q)$ are the one-loop induced mass operator and vertex
part respectively. It should be mentioned here that other
possibilities of the external Higgs attachment, to external
fermion lines or to the outer W-propagator, are not operative
due to the identity:
\be
\fr{1-\gf}{2}S_j(p)M(p)S_k(p)\fr{1+\gf}{2}\;-\;(j\leftrightarrow
k)\equiv 0.
\label{eq:ident}
\ee

The V-A character of charged currents fixes the general
structure of the mass operator before renormalization up to an
invariant function depending on $p^2$:
\be
M=\ph\fr{1-\gf}{2} f(p^2).
\ee
The on-shell renormalization with respect to quark $j$ from the
left and quark $k$ from the right introduces into the mass
operator the dependence of external masses \cite{Shab,KhP}:
\be
M_r=\tilde{f}(p^2)\ph\fr{1-\gf}{2}-f_{jk}[\ph \fr{1+\gf}{2}-
m_j\fr{1+\gf}{2}-m_k\fr{1-\gf}{2}],
\label{eq:moren}
\ee
where $f_{jk}$ and $\tilde{f}$ are expressed via the function $f$ and
masses $m_j,\;m_k$ as follows:
\be
\tilde{f}(p^2)=f(p^2)-\fr{m_j^2f_j-m_k^2f_k}{m_j^2-m_k^2},\;\;\;
f_{jk}=\fr{m_jm_k(f_j-f_k)}{m_j^2-m_k^2};\;\;
f_j=f(p^2=m_j^2).
\label{eq:f13}
\ee
For the simplicity we use the nonrenormalized
form of the mass operator and then show that the same result
remains intact for the full renormalized expression.

The expansion in $q_\mu$ is essential at the next step of the
calculation. To proceed forward with it we connect the vertex
part at zero momentum transfer, $\Gamma(p,q=0)$, to the
nonrenormalized mass operator using the Ward identity:
\be
\Gamma(p,q=0)=\fr{\partial}{\partial v}M(p).
\label{eq:Ward}
\ee
It is worth to note that the unitary gauge ensures the
cancellation of all divergencies in $\Gamma$. Moreover, this
vertex corresponds to the operator of dimension 5,
$\chi\bar{q}_j\hat{\partial}(1-\gf)q_k$, and does not
require renormalization.

The zeroth order term of the expansion of expression
(\ref{eq:struct}) in $q$ vanishes simply because it could be
reduced to the total derivative $\partial/\partial v$ from the
l.h.s. of the identity (\ref{eq:ident}). The same is true and
for any order in $\chi/v$ if we systematically
neglect the momentum associated with the Higgs field.

The first term of the expansion in $q$ does not vanish, however.
After a straightforward arithmetic we get:
\be
-i4\fr{\chi(q)}{v}\fr{(m_j^2-m_k^2)\ph(pq)}{(p^2-m_j^2)^2(p^2-m_k^2)^2}
\left(p^2\fr{\partial f}{\partial p^2}+\fr{v}{2}
\fr{\partial f}{\partial v}\right)\fr{1+\gf}{2}.
\label{eq:j-k}
\ee

The interesting feature of the formula (\ref{eq:j-k}) consists
in the vanishing of the expression in parenthesis for any given
function depending on the ratio $p^2/v^2$:
\be
\left(p^2\fr{\partial}{\partial p^2}+\fr{v}{2}\fr{\partial}{\partial v}
\right)f(p^2/v^2)\equiv 0
\ee
Despite appearance the effect is not zero due to
logarithmically divergent part of the mass operator. Taking into
account an explicit dependence of the ultraviolet cutoff
$\Lambda$ we obtain:
\be
\left(p^2\fr{\partial}{\partial p^2}+\fr{v}{2}\fr{\partial}{\partial v}
\right)\log\fr{\Lambda^2}{p^2+v^2}=-1.
\label{eq:anom}
\ee
Let us demonstrate this assertion in more details. First we
note that the GIM property makes the integral
defining function $f$ be almost convergent and therefore
be dependent on the ratio $p^2/v^2$. The word "almost" refers to
the only possible logarithmically divergent term which
originates from the longitudinal part of the W-boson propagator:
\begin{eqnarray}
f\ph\fr{1-\gf}{2}=
-i\int\fr{d^4q}{(2\pi)^4}\fr{m_t^2\qh(\ph+\qh)\qh}
{M_w^2((p+q)^2-m_t^2)(p+q)^2q^2}\fr{1-\gf}{2}\nonumber\\
=\fr{3g_w^2m_t^2}{4M_w^2}\fr{1}{16\pi^2}\log\fr{\Lambda^2}{p^2+v^2}
\ph\fr{1-\gf}{2}+...,
\label{eq:diverg}
\end{eqnarray}
where we have imposed without the lost of generality the obvious
relation $m_t^2\gg m_c^2$. In use of formulae (\ref{eq:anom})
and (\ref{eq:diverg}) the resulting expression is transformed to
the following form:
\be
-i\fr{\chi(q)}{v}\fr{(m_j^2-m_k^2)\ph(pq)}{(p^2-m_j^2)^2(p^2-m_k^2)^2}
\fr{3g_w^2m_t^2}{M_w^2}\fr{1}{16\pi^2} \fr{1+\gf}{2}.
\ee

The same answer emerges as the result of calculation with the
renormalized mass operator $M_r$. The contribution from
the counterterms in (\ref{eq:moren}) vanishes since it is
symmetric under the interchange of $m_i$ and $m_j$. We skip here
the prove of this statement which is rather simple.

To conclude this section, we have shown that the antisymmetry under
the interchange of flavors does not lead to the vanishing of
the amplitude of interest. However, strong cancellations exist
between mass operator and vertex part contributions which effectively
reduces the whole calculation to the one-loop level. The inner
loop produces just a constant multiplier proportional to the
square of the fermion Yukawa coupling on account of Eqs.
(\ref{eq:anom}) and (\ref{eq:diverg}). We performed also the
calculation in the Landau gauge, $\xi=0$, where the
answer originates from the diagrams with charged Higgs bosons.
After the complete summation over flavors the results of
calculations in different gauges coincide identically.

\section{KM predictions for the formfactors}
Using the results of the previous section it is easy to
integrate over the second loop and find corresponding
amplitudes. Now it is convenient to treat SM model and its heavy
quark extensions separately.

{\em 1. Standard Model set of flavors}\newline
\mbox{}It is easy to see that the interaction of the scalar field
with $u$ and $c$ quarks in SM is much larger than with other flavors. It is
simply explained by the $m_t^2$-enhancement factor for the
interaction with external $c$- or $u$-flavored quarks. The sum
over flavors and the loop integral are trivial so the
final answer reads as follows
\be
{\cal L}_{eff}=-\fr{3}{32\pi^4}\tilde{\delta}\fr{G_F}{\sqrt{2}}f_t^2m_s^2
\log\fr{m_b^2}{m_s^2}\;
\fr{\partial_\mu\chi}{v}\left(\bar{c}\gm\fr{1-\gf}{2}c-\bar{u}\gm
\fr{1-\gf}{2}u\right).
\label{eq:SMc}
\ee
Here we introduce the Fermi constant, $G_F=\sqrt{2}g^2/(8M^2)$,
and the Yukawa coupling of fermion $f_i=m_i/v$ in a standard way.
The integral is calculated to logarithmic accuracy and $m_s^2$
at lower limit rather symbolizes a momentum scale of order
$\Lambda_{QCD}$. Integrating by parts and applying the equation of
motion for external quarks we reduce the operator structure of
(\ref{eq:SMc}) to the common form $i\chi m_i\bar{q}_i\gf
q_i$. For down quarks (d and s) the result is proportional to
the combination $f_b^2m_c^2$. The CP-odd interaction of third
generation quarks with scalar field acquires even stronger
suppression.

The answer (\ref{eq:SMc}) is valid when the momentum transfer $q$
does not exceed $m_s$. It is clear, however, that there is an
easy way to generalize the answer at large values of $q^2$ using
$m_t^2$-dependence of the answer. The inner loop behaves as an
effective constant vertex until $|q^2|$ becomes comparable with
$m_t^2$. Therefore, the amplitude of interest may serve not only
for the scattering of quark in the scalar field background but also for
the decay of the real Higgs into the quark-antiquark pair if
scalar boson is not very heavy. The generalization at large
momentum transfers of the interaction (\ref{eq:SMc}) and of
other results from this section are accumulated in the Appendix.

Going over the calculation of the Higgs-W-W interaction we
determine first the flavor structure of the fermionic loop. Its
CP-odd part is given by the following combination:
\begin{eqnarray}
i\tilde{\delta}[d(c(b-s)t-t(b-s)c+t(b-s)u-u(b-s)t+u(b-s)c-c(b-s)u)
\nonumber\\+s(c(d-b)t-t(b-s)c+t(d-b)u-u(d-b)t+u(d-b)c-c(d-b)u)
\nonumber\\+b(c(s-d)t-t(s-d)c+t(s-d)u-u(s-d)t+u(s-d)c-c(s-d)u)]
\label{eq:fs}
\end{eqnarray}
Each product of four quark Green functions allow for the cyclic
permutation of the kind:
\[
udcs=dcsu=csud=sudc.
\]
The "degree of antisymmetry" of eq. (\ref{eq:fs}) is higher than
that of corresponding structure with external fermions (\ref{eq:u}).
This results on the stronger suppression of the interaction of W
boson with external scalar field:
\be
{\cal L}_{eff}=\fr{3}{32\pi^4}\tilde{\delta}\fr{G_F}{\sqrt{2}}
\fr{f_t^2m_c^2m_s^2}{m_w^2}
\log\fr{m_b^2}{m_s^2}\;\fr{\chi}{v}\epsilon_{\alpha\beta\mu\nu}
\partial_\alpha W_\beta\partial_\mu W^*_\nu,
\label{eq:SMW}
\ee
where $W_\beta=(\sqrt{2})^{-1}(W^1_\beta+iW^2_\beta)$. The calculation is
performed for on-shell W-bosons. The leading contribution to
(\ref{eq:SMW}) comes again from the top quark flowing inside the
inner loop at Fig. 1b.

{\em 2. KM model with the additional heavy generation(s)}\newline
\mbox{}The consideration of the SM prediction for
the CP-odd interaction with scalar field is mostly of
methodological meaning. The resulting amplitudes are too small
to produce any observable effects. Now we shall extend SM by
adding a new heavy generation with standard quantum numbers
preserving the same KM origin of CP violation. The
phenomenological constraints on the parameters of this model
are provided by the analysis of K and B meson mixing
\cite{HSS} and electroweak precision data \cite{PAS}.
When the mass of 4th generation is large and lies somewhere
between 500Gev and 1Tev one comes to the picture of the strongly
interacting Higgs-fermion sector \cite{DWTr}. We
assume that masses of h and g quarks are of order 500Gev
preserving the perturbative unitarity.

The 4 $\times$ 4 KM matrix possesses three independent CP-odd
invariants. The dynamical enhancement of flavor-diagonal
CP-violating amplitudes are associated with the invariant
corresponding to the mixing of second, third and fourth generations
of quarks \cite{HP}. The source of this enhancement is in the
change of the overall mass factor in nominators of formulae
(\ref{eq:SMc})-(\ref{eq:SMW}).

Let us demonstrate this assertion on the example of the $s$ and
$b$ quarks scattering off the Higgs background. Instead of SM
prediction with the dependence of $m_b^2m_c^2$, we may expect
the effect in the four generation model to arise with a factor
$m_g^2m_t^2$, and the total enhancement could reach $10^8$.
Large masses in nominator imply that the characteristic loop
momenta are also large.  Therefore, inside the loops, we are
legitimate to put all quark masses to zero except $m_t$, $m_g$
and $m_h$. In other words, inside the loops, we are able to
identify propagators of light quarks:
\[
c=u\equiv U;\,\,\,\, d=s=b\equiv D.
\]
After that to sufficient accuracy we derive the flavor structure of
the amplitude for the $b$ quark interaction with Higgs background
(See Ref.\cite{HP} for details):
\be
i\Ima(V^*_{ts}V_{tb}V^*_{hb}V_{hs})
b[t(g-D)h-h(g-D)t+U(g-D)t-t(g-D)U+h(g-D)U-U(g-D)h]b
\label{eq:FSfin}
\ee
The rephasing invariant combination of KM matrix elements in
(\ref{eq:FSfin}) to good accuracy coincides with that responsible
for CP-odd $B^0_S$ meson mixing. In terms of Wolfenstein
parameter it could naturally reach the order
\be
\Ima(V^*_{ts}V_{tb}V^*_{hb}V_{hs})\sim\lambda^5.
\ee

In the outer loop integral the main contribution now comes from
the longitudinal part of W-propagator. For the on-shell quarks
the result is presented again in the form of the effective
Lagrangian:
\be
{\cal L}_{eff}=\Ima(V^*_{ts}V_{tb}V^*_{hb}V_{hs})
\fr{3}{128\pi^4}f_g^2f_t^2(\log\fr{m_h^2}{m_t^2}-1)\;
\fr{\chi}{v}\left(m_b\bar{b}i\gf b-m_s\bar{s}i\gf s\right),
\label{eq:4gb}
\ee
where to sufficient accuracy we have omitted $m_w^2/m_t^2$
suppressed terms. We should take into account, of course, the
range of validity of perturbative analysis in this model. When
the Yukawa constants become sufficiently bigger than unity the perturbative
expansion does not work and we have to deal with a strong
coupling regime. In our case, however, all "dangerous" vertices are
proportional to the combination of $V_{hj}f_h$ or
$V_{jg}f_g$. According to Refs. \cite{HSS,HP} we take non
diagonal KM matrix elements not exceeding $\lambda$ and
therefore we could extend the perturbative analysis until
$f_{h(g)}\sim\lambda^{-1}$.

The interaction with W-boson in this model is also enhanced in
comparison with SM case. However, our approximation with a
complete degeneracy between light quarks inside the loops is not
operative for the interaction with W-boson because the amplitude
(\ref{eq:fs2}) vanishes in this limit. To obtain a nonvanishing
effect we must take into account the mass of $b$ quark and to that
reason it is the $m_b^2$-suppressed effect:
\be
{\cal L}_{eff}=-\Ima(V^*_{ts}V_{tb}V^*_{hb}V_{hs})
\fr{3}{32\pi^4}\fr{G_F}{\sqrt{2}}m_b^2\log\fr{m_w^2}{m_b^2}
\;\fr{\chi}{v}\epsilon_{\alpha\beta\mu\nu}
\partial_\alpha W_\beta\partial_\mu W^*_\nu.
\label{eq:4gW}
\ee

\section{Generalization at nonzero temperature and adiabatic
baryogenesis}

The interest to the CP-odd interaction of quarks and W-bosons
with a nonuniform scalar field background is mainly inspired by
the problem of systematic change of baryon number during the
electroweak phase transition. The results of previous sections
cannot be directly used in this context because they deserve a
considerable modification to meet the high temperature
conditions. Before doing that we would like to remind some
principal points of electroweak baryogenesis without going into
details and following in general the reviews \cite{Dol,D}.

It is commonly understood now that the observed ratio of baryon
to photon densities in the Universe,
\be
\fr{N_B-N_{\bar{B}}}{N_\gamma}\sim10^{-10},
\ee
could be achieved, in principle, during the
electroweak phase transition. Moreover, the SM
content of fields may generate all three Sakharov's conditions
\cite{Sakh} necessary for baryogenesis.

The first criterion of the microscopic baryon number
nonconservation is fulfilled due to the
anomaly in the current associated with this number:
\be
\partial_\mu j_b^\mu=\fr{3\al_w}{8\pi}F_{\mu\nu}^a\FD_{\mu\nu}^a.
\ee
The r.h.s. of this equation is the total derivative,
$\partial_\mu K_\mu$, and
it implies that baryon number conserves in the specific
combination with the topological charge, $\Delta(Q_b-Q_{cs})=0$,
where by definition
\be
Q_{cs}=\int d^3xK_0=\fr{g^2}{16\pi^2}\epsilon^{ijk}\int
Tr\left(F_{ij}W_k+\fr{2}{3}igW_iW_jW_k
\right)d^3x.
\label{eq:Qcs}
\ee
Here $W_i=W_i^a\tau_a/2$ and
$F_{ij}=\partial_iW_j-\partial_jW_i+ig[W_i,W_j]$. At zero
temperatures the effects of tunneling between topologically
distinct vacua are exponentially suppressed. In contrast to that
at very high temperature in the unbroken phase the exponential
suppression is removed and the rate of the processes with
$\Delta Q_{cs}=\Delta Q_b \neq 0$ is believed to go as
\be
\Gamma=c\al_w^4T^4,
\ee
where $c$ is some dimensionless unknown coefficient. With the
growth of the scalar field v.e.v. the exponential suppression is
switch on at some value $v_0$ which is not known to sufficient
accuracy. Simple arguments suggest the order of magnitude
estimate for this value: $g_wv_0\sim
\al_w T$, whereas the semiclassical analysis of sphaleron
processes gives a numerical enhancement for this value:
$g_wv_0\sim 14\al_w T$ \cite{AMcL}.

The second requirement could also be satisfied. The
analysis of the effective potential at the critical point
suggests the possibility of the first order phase transition
between symmetric, $v=0$, and broken, $v=v_c\neq0$, phases. The
propagation of the domain walls separating two phases through
the relativistic plasma breaks the thermal equilibrium and
generates the arrow in time. After the transition all processes
with $\Delta Q_b\neq 0$ should be suppressed, i.e. $v_c>v_0$, to avoid
the baryon number erasure \cite{Rub}. The C and CP violation shifts the
processes with $\Delta Q_b\neq0$ toward a preferential
production of baryons.  However, the amount of CP violation
inherently presented in SM could lead to an asymmetry,
according to Refs.\cite{Gav,HS}, of order $n/s\sim 10^{-27}$ as
best and this is the main obstacle for SM explanation of the
baryon number of the Universe. In what follows we concentrate
ourself on the analysis of CP violation developed in KM model
with different numbers of generations assuming that all other
conditions of baryoproduction are indeed satisfied.

When the domain wall is thick one comes to the "adiabatic"
treatment of baryogenesis \cite{DHSS,McLSTV,CKN1}. In this case,
considerably simplified in comparison with the generic
situation, the slowly varying in time vacuum expectation value of
the scalar field serves as a chemical potential for the
Chern-Simons charge:
\be
L_{int}=A\dot{v}Q_{cs}
\label{eq:op}
\ee
The appropriate sign of A and B generates the arrow for the
sphaleron processes toward the observed density of baryons.
Then the total amount of baryons could be estimated as follows
\cite{D}:
\be
Q_b\sim\fr{1}{T}\int \Gamma A\dot{v}\,dt
\ee

There are two different scheme to generate CP-violating
operators analogous to (\ref{eq:op}) in the model. First one
refers to the case of small vacuum expectation values of the
scalar field and could be achieved through the many loop
mechanism like that proposed in different context by Ellis and
Gaillard \cite{EG} and readdressed for baryogenesis by
Shaposhnikov \cite{Shap} (See also the review \cite{D}). The growth
of v.e.v. suggests the switch to another regime \cite{Shap}
where the low-loop amplitudes are less suppressed.  We estimate
the critical field $v_1$ where the change of these regimes
occurs as following:
\be
\fr{v_1^2}{(\pi T)^2}\sim \fr{1}{16\pi^2}
\ee
The v.e.v. of the scalar field is normalized at $(\pi T)^2$
as it follows from the finite temperature Feynman rules.
It is quite possible also that $v_1<v_0$ i.e. this value
develops {\em after} all sphaleron-like processes become
suppressed and therefore the two-loop induced interaction does
not affect the baryon density. The opposite case of $v_1>v_0$
deserves special consideration.

{\em 1. Standard Model set of flavors}\newline
\mbox{}If the baryon number violating processes shut down at
sufficiently large v.e.v. it is reasonable to generalize the
two loop mechanism described in Sections 2 and 3 at nonzero
temperatures. First, we note that in any case the result
expected is really tiny and there is no need in the exact
calculation. The calculation itself is now more complicated than
in the case of $T=0$ due to the existence of another
dimensional parameter $T$ and the lack of Lorentz invariance.
Thus, simple arguments leading to the elimination of the
momentum dependence come from the inner loop do not work and we have
to deal with a real two-loop calculation.

Let us integrate over all quark fields. Then the effective
operators governing baryoproduction could be composed from
neutral scalar and vector boson fields. It is clear that in
leadin order they originate from the two loop diagram at Fig.1b.
The quark operators contribute to the effect at the next
three-loop order and thus are neglected.

The quark mass dependence of the answer arises through the mass
insertions \cite{Shap} affecting the stronger compensation
of different diagrams than it happens at $T=0$. At the first
glance the flavor structure (\ref{eq:fs}) of the fermionic loop
implies that the effect arises first in the $v^{10}$-order
together with antisymmetric product of Yukawa couplings:
\be
(f_t^2-f_c^2)(f_t^2-f_u^2)(f_c^2-f_u^2)
(f_b^2-f_s^2)(f_b^2-f_d^2)(f_s^2-f_d^2)\simeq f_t^4f_c^2f_b^4f_s^2
\label{eq:Yuk}
\ee
Two powers of Yukawa coupling originate here from the vertices
with charged Higgs and are not accompanied by $v^2$.
However, the actual degree of suppression is even stronger.
The expression (\ref{eq:fs}) changes the sign under the
permutation of up and down families of flavors whereas
the lowest order dependence of Yukawa couplings (\ref{eq:Yuk})
is explicitly symmetric. It means that in this order the
diagrams with up quarks flowing inside mass operator cancel
those with down quarks. To avoid this cancellation one has to
introduce the weak isospin asymmetry
via an additional loop with the exchange of U(1) gauge
boson or via two additional mass insertions for
the top flavor. The second possibility gives bigger interaction
in the chosen conditions when the tree level dominates over loop
corrections. Thus, the resulting estimate takes the form:
\be
L_{int}\sim\tilde{\delta} \fr{3}{8\pi^2}f_t^6f_c^2f_b^4f_s^2
\fr{\dot{v}v^{11}}{(\pi T)^{12}}
\fr{g^2}{16\pi^2}\epsilon^{ijk}\int (W^{(1)}_i\partial_jW^{(1)}_k+
W^{(2)}_i\partial_jW^{(2)}_k)d^3x
\label{eq:TSM}
\ee

This interaction does not literally correspond to the
form (\ref{eq:op}). We see that it depends on
the global orientation $n_a$ of the vacuum configuration of the
scalar field doublet $\phi$:
\be
\partial_\alpha W^1_\beta\partial_\mu W^1_\nu+
\partial_\alpha W^2_\beta\partial_\mu W^2_\nu=
\partial_\alpha W^a_\beta\partial_\mu W^a_\nu-
n^an^b\partial_\alpha W^a_\beta\partial_\mu W^b_\nu
\ee
where
\[
v^2n^a\equiv\phi^\dagger\tau^a\phi.
\]
We would assume that the nucleation of the new phase
occurs with random orientation of $n^a$ in different bubbles. Simple
average over this orientation gives the coefficient $2/3$. The
absence of terms trilinear in gauge field in the expression
(\ref{eq:TSM}) does not mean that they cannot be induced or do not
affect sphaleron-like processes. Since the $SU(2)\times U(1)$
symmetry is spontaneously broken we could regard quadratic and
trilinear terms as independent operators. It
is clear that any nonorthogonal to $Q_{cs}$ linear
combination of these operators generates an effective
chemical potential for $Q_b$. Combining several factors we estimate
the resulting asymmetry to arise at the level:
\be
\fr{N_B-N_{\bar{B}}}{N_\gamma}\sim 10^{-2}\al_w^4
\fr{3}{8\pi^2}\tilde{\delta}f_t^6f_c^2f_b^4f_s^2
\fr{v^{12}}{(\pi T)^{12}}
\ee
Numerically this asymmetry is really tiny and reaches $10^{-40}$
in the most optimistic assumptions about the sphaleron cuttoff.

{\em 2. KM model with the additional heavy generation(s)}\newline
\mbox{}
The estimate of the effect in this case goes along the
same way if we take the unknown heavy masses somewhere around
the mass of the top quark, $m_{h(g)}\sim m_t$. Thus, simply renaming
quarks and phases in the expression (\ref{eq:TSM}) and taking
into account the isospin splitting inside third generation we obtain:
\be
L_{int}\sim\Ima(V^*_{ts}V_{tb}V^*_{hb}V_{hs})\fr{3}{8\pi^2}f_h^4f_t^4
f_g^4f_b^2\fr{\dot{v}v^{11}}{(\pi T)^{12}}\fr{g^2}{16\pi^2}
\epsilon^{ijk}\int (W^{(1)}_i\partial_jW^{(1)}_k+
W^{(2)}_i\partial_jW^{(2)}_k)d^3x.
\ee

The case of very heavy $h$ and $g$ quarks, $m^2_{h(g)}\gg m_t^2$,
requires separate treatment because at some v.e.v. $v_2$ their
masses become comparable with temperature, $f_hv_2\sim T$, and
cannot be treated as mass insertions. The simplification of the
calculation is possible if we take $f_hv\gg T$. It allows to use
zero temperature expressions for the inner loop with quarks
from the fourth generation and calculate the rest of diagram
with "light quarks" in accordance with the finite temperature
technique. After the straightforward calculation we get:
\begin{eqnarray}
L_{int}=\Ima(V^*_{ts}V_{tb}V^*_{hb}V_{hs})\fr{3}{32\pi^2}
\lambda(5)(f_h^2-f_g^2)f_t^2
f_b^2\fr{\dot{v}v^{3}}{(\pi T)^{4}}\fr{g^2}{16\pi^2}\times\nonumber\\
\epsilon^{ijk}\int (W^{(1)}_i\partial_jW^{(1)}_k+
W^{(2)}_i\partial_jW^{(2)}_k)d^3x,
\label{eq:T4g}
\end{eqnarray}
where $\lambda(5)=\fr{31}{32}\zeta(5)\simeq 1$.

It is instructive to compare the size of this interaction with the
CP-even coupling of the topological charge and the relative
phase of different scalars which arises in the multi Higgs
models \cite{McLSTV}. This effective interaction governs
baryoproduction in the model when the CP-violating phase in
Higgs sector changes
from zero to some finite value $\theta$. Being generated at
one-loop level, this coupling is proportional to $m_t^2\theta$
and this is again $v^4$-dependence \cite{D}. We would like to stress
that the relative smallness of the effect in KM model with four
generation of fermions results from: \newline {\em a)} Flavor
symmetry of the amplitude and corresponding $f_b^2$-suppression;
\newline {\em b)} Smallness of the CP-odd angle invariant
$\Ima(V^*_{ts}V_{tb}V^*_{hb}V_{hs})$. It could be restricted from the
data on neutral $B_S$ meson mixing and is unlikely to exceed
$\lambda^5$ in terms of Wolfenstein parameter;\newline {\em c)}
Additional limits on the mass difference of heavy quarks come
from electroweak $\rho$ parameter. Both $f_h^2$ and $f_g^2$
could be rather large. However, the analysis of electroweak
precision data implies that $h$ and $g$ quarks must be sufficiently
degenerate in masses, $(f_h-f_g)^2\ll 1$. So, the
factor $f_h^2-f_g^2\simeq 2f_h(f_h-f_g)$ does not exceed unity
if we take $m_{h(g)}$ somewhere in the interval 500 Gev - 1 Tev.

If we would introduce into the theory two or more additional heavy
generations, (h, g); (h', g');..., with a large mixing between
them we would remove $f_b^2$-dependence. At the same time there is no
strict limits on the mixing between heavy generations and we
could expect the corresponding CP-odd combination of mixing
angles $\Ima(V^*_{tg'}V_{tb}V^*_{hb}V_{hg'})$ to be large.  The
third factor of suppression related to the antisymmetry under the
interchange of U and D types of quarks is still held.  Now the
CP violation occurs entirely in fermion-Higgs sector of the
theory and it is clear that potentially the KM type of models
with two or more heavy generations of quarks are capable to
produce the amount of CP-violation comparable with that of
multi-Higgs models. Unfortunately, the simplest variant of this
model is already excluded by electroweak precision data analysis.

\section{Conclusions}
We have shown that the two-loop level is sufficient to
to induce the CP-odd interaction of particles with the
scalar field background in the KM model. Together with the
magnetic quadrupole moment of the W-boson \cite{KhP1} these are
the only known examples of flavor-conserving CP-odd operators of
low dimension, dim$\leq$6, which do not vanish to two-loop
approximation in this model in zero temperature limit. The sum
over flavors inside the loops leads to the remarkable
compensation of different contributions which affects in its
turn on the effective simplification of the whole calculation.
The inner loop gives just a constant multiplier proportional to
the square of the Yukawa coupling of the fermion flowing inside
this loop. This is an explicit example of the nondecoupling of
heavy fermions in the electroweak theory: for the SM set of
flavors the result is proportional to $f_t^2$, for its four
generation extension it is $f_{h(g)}^2$, etc. The absence of
decoupling allows to extend the calculation at large momentum
transfers and keep the effective one-loop level of difficulty.
The resulting amplitudes describes the CP violation in the decay
of the Higgs boson to accuracy ${\cal O}(m_{Higgs}^2/m_f^2)$,
where $m_f$ is the mass of the heaviest quark.

The attempt to plug this interaction into the scheme of
electroweak baryogenesis at the temperature of the phase
transition faces with the poor knowledge of the vacuum
expectation value $v$ at which all transitions with $\Delta Q_b\neq0$
become suppressed. In the assumption that this value is rather
large, $g_w v$ is of order several units of $\al_wT$, the two-loop
CP-violating mechanism dominates over multi-loop ones. The
prescriptions of the finite temperature diagram technique
together with the flavor symmetry of diagrams provide a strong
suppression of the interaction of interest. From naive
expectations in the SM case the effect is proportional to the
minimal CP-odd combination of mixing angles and Yukawa
couplings $\tilde{\delta}f_t^4f_c^2f_b^4f_s^2v^{10}(\pi T)^{-10}$. The
additional antisymmetry with respect to interchange of U and D
types of quarks makes the total degree of suppression be even
smaller: $\tilde{\delta}f_t^6f_c^2f_b^4f_s v^{12}(\pi T)^{-12}$.
The analysis of the KM model with four generations of quarks is
performed in the most interesting situation when the heaviest
quark masses are comparable with temperature. The corresponding
size of the effect now is: $\Ima(V^*_{ts}V_{tb}V^*_{hb}V_{hs})
(f_h^2-f_g^2)f_t^2f_b^2v^{4}(\pi T)^{-4}$.

All three factors of suppression pointed out in the previous
section will be held in a generic nonadiabatic case.
The numerical raise of the effect in the presence of the narrow
wall is connected mainly with the change of temperature infrared
cutoff. Instead of powers of $\pi T$ in the denominator of
formulae (\ref{eq:TSM})-(\ref{eq:T4g}) one could expect the
normalization on parameters characterizing the propagation and
collisions of quasiparticles in the hot plasma $\sim g_sT$.

Our intention to make KM type of CP violation useful for
electroweak baryogenesis implies to introduce two new
heavy generations at least. The amount of
CP violation developed at high temperatures in this model does
not differ considerably from that of very popular multi-Higgs
extensions of SM. To make this model be consistent with electroweak
precision data constraints on isospin symmetric observables \cite{PAS}
one has to introduce
additional bosons into the theory to compensate the large positive
contribution to the S-parameter.
Another weak point of the proposed analysis is in the neglection
of other requirements needed for baryoproduction. It is clear
that heavy fermions affect the character of the phase
transition. In the perturbative treatment of the effective
potential they weaken the first order phase transition. To
avoid the baryon number erasure after the transition and ensure
the $T=0$ Higgs boson mass to satisfy modern experimental limits
one comes again to the necessity of additional bosons in the theory
\cite{AH}. This would lead in a generic situation to new
sources of CP violation besides complexity of the KM matrix.

\begin{flushleft}{\Large\bf Acknowledgments}\end{flushleft}
I thank V. Chernyak, C. Hamzaoui, I. Khriplovich, and A.
Yelkhovsky for helpful discussions. This work was
supported in part by the I.C.F.P.M., INTAS grant \#\,93-2492.

\begin{flushleft}APPENDIX\end{flushleft}
The generalization of static formfactors at large momentum
transfers is performed below in use of the $f^2$ dependence of
the one-loop Higgs-fermion-fermion vertex. The
characteristic momentum inside this loop is large and
determined by the $m^2$, the mass of the heaviest fermion. This
allows us to take this vertex in the form:
\be
M=\fr{\chi}{v}\fr{3f^2}{8\pi^2}(\ph_1+\ph_2)\fr{1-\gf}{2}+
{\cal O} (\fr{m_w^2}{m^2};\fr{q^2}{m^2};\fr{p_1^2}{m^2};
\fr{p_2^2}{m^2}),
\label{eq:vert}
\ee
where we used again the unitary gauge. Besides the trivial
kinematic factor $\ph_1+\ph_2$ of incoming and outgoing fermion
momenta this vertex is momentum
independent and therefore the whole computation is effectively of
one-loop difficulty level.

In the SM the growth of $|q^2|$ brings the additional
serious suppression of the interaction of interest. Now the
CP-violating amplitude reads as follows:
\begin {eqnarray}
M=-\fr{3}{32\pi^4}f_t^2\fr{G_F}{\sqrt{2}}\fr{m_s^2m_b^4}{q^4}
\log\fr{|q^2|}{m_b^2}\;
\fr{\chi}{v}(m_c\bar{c}i\gf c-m_u\bar{u}i\gf u)\;\;\;\;{\rm at}\;\;\;
m_b^2\ll |q^2|\ll m_w^2;\nonumber\\
M\simeq\fr{9}{128\pi^4}f_t^2\fr{G_F}{\sqrt{2}}\fr{m_s^2m_b^4}{q^2m_w^2}
\log\fr{m_w^2}{m_b^2}\;
\fr{\chi}{v}(m_c\bar{c}i\gf c-m_u\bar{u}i\gf u)
\;\;\;\;{\rm at}\;\;\;m_w^2\ll |q^2|\ll m_t^2,
\end{eqnarray}
where we hold only $\log m_b^2$-contributions, i.e. infrared
enhanced terms. We see the restoration of the factor $m_b^4$, not
unlike it happens at high temperature. We skip here the
calculation of the interpolation between these two formulae at
$q^2\sim m_w^2$ which is also, of course, very simple.

The W-boson interaction with
Higgs drops with the growth of $q^2$ as follows:
\begin{eqnarray}
M\simeq\fr{3}{32\pi^4}\tilde{\delta}\fr{G_F}{\sqrt{2}}
\fr{f_t^2m_c^2m_b^4m_s^2}{q^4m_w^2}
\log\fr{|q^2|}{m_w^2}\;\fr{\chi}{v}\epsilon_{\alpha\beta\mu\nu}
\partial_\alpha W_\beta\partial_\mu W^*_\nu,\nonumber
\\ \;\;\;\;{\rm at}\;\;\;m_w^2\ll |q^2|\ll m_t^2,
\end{eqnarray}
where from the reasons of simplicity we hold only contributions
proportional to $\log m_w^2$.

The four generation case allows for the generalization at larger
momenta. For the physically interesting scale relation $m_w^2\ll
|q^2|\sim m_t^2\sim|q^2-m_t^2|\ll m_{h(g)}^2$ we obtain the
following result:
\begin{eqnarray}
M=\Ima(V^*_{ts}V_{tb}V^*_{hb}V_{hs})
\fr{3}{128\pi^4}f_g^2f_t^2(\log\fr{m_h^2}{m_t^2}+
\fr{m_t^2-q^2}{q^2}\log\fr{m^2}{|m^2-q^2|}-2)\times\nonumber\\
\fr{\chi}{v}\left(m_b\bar{b}i\gf b-m_s\bar{s}i\gf s\right),
\end{eqnarray}
We see that the ratio of the CP-odd couplings to CP-even ones for
$s$ and $b$
flavors is parametrically suppressed in fact only by the
combination of mixing angles and does not drop with the growth
of $q^2$. Unfortunately, the size of this amplitude, being
considerably enhanced in comparison with SM predictions, is not
likely to be observed mainly because of the smallness of mixing
angle combination. Finally, the CP-odd interaction of W-boson with Higgs
reads as follows:
\begin{eqnarray}
M\simeq-\Ima(V^*_{ts}V_{tb}V^*_{hb}V_{hs})
\fr{3}{32\pi^4}\fr{G_F}{\sqrt{2}}\fr{f_h^2m_b^2m_w^2}{q^2}
\log\fr{|q^2|}{m_w^2}\left(2+\fr{m_t^2}{q^2}\log\fr{|q^2-m_t^2|}{m_t^2}
\right)\times\nonumber\\ \fr{\chi}{v}\epsilon_{\alpha\beta\mu\nu}
\partial_\alpha W_\beta\partial_\mu W^*_\nu.
\label{eq:A4gW}
\end{eqnarray}

\end{document}